# Acoustic realization of projective mirror Chern insulators

Tianzi Li, Luohong Liu, Qicheng Zhang, and Chunyin Qiu[*]

Key Laboratory of Artificial Micro- and Nano-Structures of Ministry of Education and School of Physics and Technology, Wuhan University, Wuhan 430072, China

[*] To whom correspondence should be addressed: cyqiu@whu.edu.cn

**Abstract.** **Symmetry plays a key role in classifying topological phases. Recent theory shows that in the presence of gauge fields, the algebraic structure of crystalline symmetries needs to be projectively represented, which enables unprecedented topological band physics. Here, we report a concrete acoustic realization of mirror Chern insulators by exploiting the concept of projective symmetry. More specifically, we introduce a simple but universal recipe for constructing projective mirror symmetry, and conceive a minimal model for achieving the projective symmetry-enriched mirror Chern insulators. Based on our selective-excitation measurements, we demonstrate unambiguously the projective mirror eigenvalue-locked topological nature of the bulk states and associated chiral edge states. More importantly, we extract the non-abelian Berry curvature and identify the mirror Chern number directly, as conclusive experimental evidence for this exotic topological phase. All experimental results agree well with the theoretical predictions. Our findings will shine new light on the topological systems equipped with gauge fields.**

The discovery of topological insulators has led to an upsurge of exploring new materials with nontrivial topological properties[1-3]. After a revolutionary topological classification by internal symmetries[4,5] (including time-reversal, particle-hole, and/or chiral symmetries), the categorization has been generalized to the spatial symmetries of crystalline systems[6-16]. In particular, the introduction of nonsymmorphic symmetries, which combine a point-group operation and a fractional lattice translation, greatly enriches the classification of topological materials[17-20]. The uniqueness of a nonsymmorphic space group lies in its algebraic structure of group representations[21]. Comparing to that of symmorphic symmetry, the multiplication rule of the point-group representation features an additional phase factor and enforces a projective representation for nonsymmorphic symmetry. Very recently, it unveils that in the presence of gauge degrees of freedom, spatial symmetries of a crystalline system need to be projectively represented, which completely changes the fundamental algebraic structure of the original symmetry group[22-27]. This inspires to establish a systematic projective topological classification based on the extraordinarily rich interplay between all variety of gauge and crystalline symmetries, particularly in various artificial systems[28-35] with abundant gauge symmetries[36-39].

Here, we propose a general recipe for constructing projective mirror symmetry and report an experimental realization of mirror Chern insulators (MCIs) in spinless systems. As the earliest discovered crystalline topological phase[8-11], the MCI can be viewed as two copies of mirror symmetry-connected Chern insulators under time-reversal invariance. Since the time-reversal symmetry maps two mirror eigenspaces with each other, the MCI requires spinful mirror symmetry with eigenvalues $m_z = \pm i$, which is naturally absent in the spinless classical artificial structures[28-35]. Therefore, it seems impossible to achieve MCIs in such spinless artificial systems. Recently, it has been demonstrated that one can realize spinless MCIs by constructing bilayer-twisted Hofstadter models[40,41]. In this work, we introduce a general approach to construct projective mirror symmetry and associated MCIs in Bosonic systems. More specifically, we introduce a homobilayer lattice model with pairwise positive-negative interlayer hoppings, where the mirror operator can be projectively represented under gauge transformation. Intriguingly, the projective mirror symmetry behaves like spinful although the bilayer system is spinless in its nature. Furthermore, we conceive a minimal model to achieve projective mirror symmetry-protected MCIs, and demonstrate it unambiguously with acoustic experiments. Based on our elaborately-developed experimental technique, we can distinguish the bulk (or edge) states into different mirror subspaces, and explore the topological physics at an ultimate level of wavefunctions. We identify this new topological phase not only by measuring its bulk and edge spectra, but also by directly characterizing its mirror Chern number (MCN). The latter, never reported previously, serves as a conclusive manifestation for the unique band topology of the projective MCI (PMCI). Our findings can spur on new activities to construct novel topological phases with projective symmetries.

**Results**

**Projective mirror symmetry and a minimal model of PMCI**

We introduce first a generic route for constructing projective mirror symmetry. As sketched in Fig. 1a, we start with a bilayer system of Hamiltonian $H = \begin{pmatrix} h & T \\ T^\dagger & h \end{pmatrix}$, which consists of two *arbitrary* identical monolayers with Hamiltonian $h$. Mathematically, if the interlayer coupling $T = T^\dagger$, this homobilayer Hamiltonian $H$ will exhibit a conventional basal mirror symmetry $M_z = \sigma_1 \mathbb{I}$ with $M_z^2 = 1$, where $\sigma_1$ (or $\sigma_i$ and $\tau_i$ below) denotes a Pauli matrix and $\mathbb{I}$ is an identity matrix of the same dimension as $h$. By contrary, if $T$ satisfies $T = -T^\dagger$, the homobilayer system itself does *not* respect $M_z$ since the conventional mirror operation switches the sign of the interlayer coupling. However, the system satisfies projective mirror symmetry $\mathcal{M}_z = GM_z = i\sigma_2 \mathbb{I}$, where the gauge transformation $G = \sigma_3 \mathbb{I}$ switches the sign of the interlayer coupling again. More intuitively, figure 1a illustrates the key difference of the interlayer couplings between the systems with $M_z$ and $\mathcal{M}_z$. In contrast to the former with mirror-symmetric interlayer hoppings (gray bonds), the latter features pairwise positive (red bond) and negative (blue bond) interlayer hoppings of the same amplitudes. Note that the interlayer hoppings refrain from being nearest-neighboring ones, since such Hamiltonian components respect $M_z$ symmetry. Intriguingly, although the model proposed is spinless in its nature, the projective mirror symmetry $\mathcal{M}_z$ behaves like spinful with $\mathcal{M}_z^2 = -1$, which is essential for realizing the PMCI below.

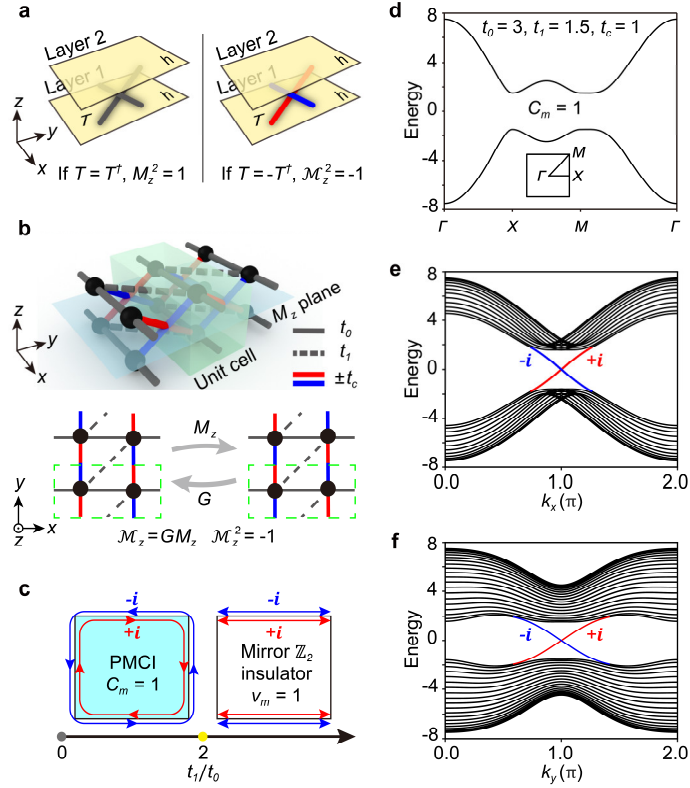

**Fig. 1 | 2D PMCI from projective mirror symmetry. a**, Schematics of a homobilayer system (left) with conventional mirror symmetry ($M_z$) and a homobilayer system (right) with projective mirror symmetry ($\mathcal{M}_z$). In contrast to the former with mirror-symmetric interlayer hoppings (gray bonds), the latter features pairwise positive (red bond) and negative (blue bond) interlayer hoppings of the same amplitudes. **b**, A minimal four-band PMCI model. Top: detailed lattice structure. Bottom: a top view of the lattice model, where the green dashed rectangle indicates a unit cell. Importantly, the sign flip of the interlayer hoppings by $M_z$ is switched back by the gauge transformation $G$. **c**, Phase diagram plotted with $t_1/t_0$, manifesting two nontrivial phases of different topological indices. The color arrows indicate the propagations of edge states carrying different projective mirror eigenvalues $m_z = \pm i$. **d**, Bulk band structure exemplified for a PMCI. Each band is two-fold degenerate everywhere. **e,f**, The associated ribbon spectra along the $x$ and $y$ directions, respectively, where $\pm i$ distinguish the chiral edge states illustrated in **c**.

To construct a PMCI, we consider a four-band homobilayer model of rectangular lattice (Fig. 1b). It is a minimal model that allows the presence of two gapped bands in each mirror subspace. Notably, the monolayers are coupled with a pair of next-nearest-neighboring interlayer hoppings of *opposite* signs (i.e., $\pm t_c$), as a key feature for the presence of projective mirror symmetry. More specifically, the monolayer Hamiltonian reads $h = t_0[(1 + \cos k_x)\tau_1 + \sin k_x \tau_2] + t_1(\cos k_y \tau_1 - \sin k_y \tau_2)$, where $k_x$ and $k_y$ are lattice momenta in the $x$ and $y$ directions, $t_0$ and $t_1$ (both assumed to be positive) are nearest- and next-nearest-neighboring intralayer hoppings, respectively, and the interlayer coupling can be written as $T = 2it_c \sin k_y \tau_3$. As illustrated in the bottom panel of Fig. 1b, the homobilayer Hamiltonian $H = \sigma_0 h + i\sigma_2 T$ respects the projective mirror symmetry $\mathcal{M}_z = i\sigma_2 \tau_0$, since the sign flip of the interlayer hoppings by $M_z = \sigma_1 \tau_0$ is switched back by the gauge transformation $G = \sigma_3 \tau_0$. Besides, this system also exhibits a conventional twofold rotation $C_{2z} = \sigma_0 \tau_1$ and a projective inversion $\mathcal{P} = GP = i\sigma_2 \tau_1$. In particular, a combination of the projective inversion $\mathcal{P}$ and the time-reversal symmetry $\mathbb{T}$ returns any momentum to itself, which satisfies $(\mathcal{P}\mathbb{T})^2 = -1$ and thus enforces the bulk bands doubly degenerate over the Brillouin zone (BZ). A complete symmetry analysis of this model can be seen in Supplementary Notes 1. Notably, as sketched in Fig. 1c, the system belongs to a PMCI phase if the dimensionless intralayer hopping parameter $t_1/t_0 < 2$ (see Supplementary Note 2). It features a topological band gap with nontrivial MCN $C_M = (C_{+i} - C_{-i})/2 = 1$, where $C_{\pm i} = \pm 1$ are Chern numbers calculated for the two-band subspaces of mirror eigenvalues $m_z = \pm i$. (Note that the total Chern number $C_{+i} + C_{-i} = 0$ vanishes due to the presence of time-reversal symmetry.) This implies that any truncated sample boundary supports one pair of counter-propagated and $m_z$-locked chiral edge states. The above physics can be seen in Figs. 1d-1f by a concrete example.

Some remarks are provided for our lattice model. (**i**) The phase diagram in Fig. 1c is irrelevant to the strength of the *interlayer* coupling $t_c$. The bulk gap closes if any of the three hoppings (i.e., $t_0$, $t_1$ and $t_c$) vanishes, and the MCN transitions into $-1$ if one of them changes its sign. (**ii**) The band gap closes and reopens as the continuous growth of $t_1/t_0$, leading to a topological transition at $t_1/t_0 = 2$ (Fig. 1c) When $t_1/t_0 > 2$, the system has a zero MCN but a nontrivial mirror $\mathbb{Z}_2$ number. Unlike the PMCI phase, the mirror $\mathbb{Z}_2$ insulator features a pair of helical edge states on the $x$-directed edges. Below we focus on the MCI phase although the novel mirror $\mathbb{Z}_2$ insulator is also very interesting in its own right (see Supplementary Fig. 1). (**iii**) The projective inversion or twofold rotation is not necessary for constructing a PMCI.

A perturbation that breaks both symmetries will not influence the band topology of the system (see Supplementary Note 3 and Supplementary Fig. 2). (**iv**) Here we only demonstrate a minimal four-band model with $C_M = 1$. The design can be extended to realize PMCIs of higher MCNs (see Supplementary Note 4 and Supplementary Fig. 3).

**Acoustic realization of a 2D PMCI**

As shown in Fig. 2a, the tight-binding model in Fig. 1b can be directly implemented with acoustic cavity-tube structures. Physically, as visualized more clearly in the zoomed-in structure, the cavity resonators emulate atomic *p*-orbitals and the narrow tubes introduce hoppings between them[42-44]. With structure details provided in Supplementary Note 5 and Supplementary Fig. 4, our acoustic PMCI features an effective onsite energy ≈5700 Hz and effective hoppings $t_0 \approx 71$ Hz, $t_1 \approx 60$ Hz, and $t_c \approx 40$ Hz. The experimental sample, consisting of $14 \times 14$ unit cells in the *x* and *y* directions, was 3D-printed by a photosensitive resin material at a fabrication error of ~0.1 mm. To experimentally excite and detect the acoustic information, small holes were perforated on the cavity resonators for inserting the sound source or probe, which were sealed when not in use.

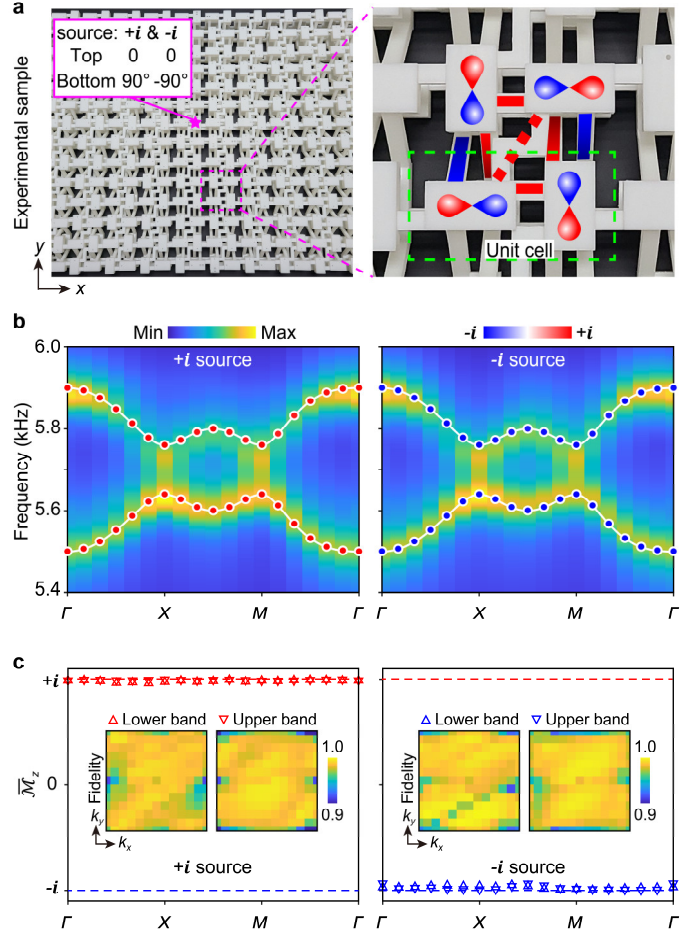

**Fig. 2 | Acoustic realization of a 2D PMCI and its bulk characterization. a**, Experimental sample, where the zoomed-in structure illustrates a detailed correspondence to the lattice model. To selectively excite the bulk bands of $\pm i$ subspaces, two point-like sound sources (magenta star) with specified phase information are separately located at middle of the two layers. **b**, Bulk spectra (color scale) excited by the $\pm i$-sources, compared with their theoretical band structures (white lines). **c**, Expectation values ($\bar{\mathcal{M}}_z$) of the projective mirror operator extracted for both bands, which are encoded to the bulk bands in **b** for clarity (see colored dots). The insets exhibit high fidelities (~98% on average) of the selectively-excited wavefunctions over the first BZ.

## Characterizing the bulk spectra and projective algebra

First, we experimentally characterized the bulk property of our acoustic PMCI. To distinguish the bulk information in each mirror subspace, selective excitation was further developed from the experimental technique in Ref. 45. Specifically, we placed a pair of broadband point-like sound sources in the middle of the sample, where the bottom-layer one carries a phase delay of $\pm\pi/2$ with respect to the top-layer one (Fig. 2a, inset). The sources will selectively excite the subspaces of projective mirror eigenvalues $m_z = \pm i$, according to the symmetry characteristics of the eigenstates. For each type of excitation, we scanned the acoustic response over the sample, and performed Fourier transform to achieve the frequency spectrum in momentum space. The experimental data (color scale) in Fig. 2b capture well the theoretical band structures (white lines). The band broadening in the experimental data is mainly caused by finite-size effect and unavoidable acoustic dissipation. To further identify the selectively-excited bulk states, we extracted the vectorial Green function in frequency-momentum space and then obtained the normalized wavefunctions $u_k(\omega_k)$ at their predicted eigenfrequencies (see Supplementary Note 6 and Supplementary Fig. 5). Using the measured wavefunctions, we computed the expectation values of the projective mirror operator, $\bar{\mathcal{M}}_z = \langle u_k|\mathcal{M}_z|u_k\rangle$. As shown in Fig. 2c, the experimental results are very close to $\pm i$ (at an average relative error of ~3.0%), as a consequence of the high fidelities (insets) of the selectively-excited wavefunctions. (The fidelity is defined by a projection of the measured wavefunction to the theoretical one.) The above fact clearly evidences that the bulk spectra in Fig. 2b indeed belong to the $m_z = \pm i$ mirror subspaces separately, which serves as a hallmark manifestation of the distinctive projective algebra $\mathcal{M}_z^2 = -1$ and $(\mathcal{P}\mathbb{T})^2 = -1$ in our acoustic PMCI. For clarity, we encode the measured mirror expectation values to the bulk bands in Fig. 2b (colored dots).

**Characterization of the topological invariant**
Next, we checked the unique bulk topology of our acoustic PMCI by directly measuring the topological invariant (i.e., MCN). It is well accepted that experimentally detecting topological invariant is always extremely challenging since it demands a precise measurement of wavefunctions. It becomes even more difficult to characterize the MCN here, because we must further distinguish the wavefunctions in two energetically degenerate mirror subspaces. To the best of our knowledge, an experimental characterization of MCN has not yet been reported elsewhere since the discovery of the spinful MCIs one decade ago[8-11]. Thanks to our skillful selective-excitation technique, we can directly calculate the Chern number of the lower band for each mirror subspace, based on our experimentally extracted bulk wavefunctions $u_k$.

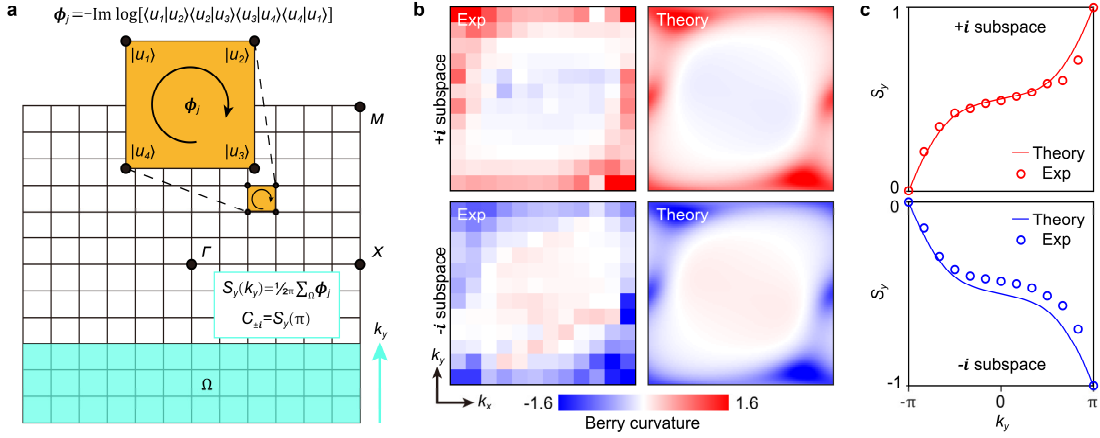

**Fig. 3 | Experimental identification of the MCN $C_M = 1$. a**, Discretized first BZ and Berry flux $\phi_j$ defined for each plaquette $j$. The Chern number $C_{\pm i}$ can be calculated by accumulating $\phi_j$ row by row for each mirror subspace, as illustrated by the cyan area. **b**, Berry curvature distributions extracted for the lower bands of $\pm i$ subspaces, together with those theoretical predictions for comparison. **c**, $k_y$-evolved Berry flux accumulation $S_y(k_y)$ defined in **a**, where $S_y(\pi) = \frac{1}{2\pi}\sum_{BZ} \phi_j = \pm 1$ gives the Chern number $C_{\pm i} = \pm 1$ for the corresponding mirror subspace.

Notice that the momentum space is discretized for our finite-sized sample (Fig. 3a). On the discretized BZ, the Berry curvature (defined by the Berry flux per plaquette) can be linked to the Wilson loop by Stokes' theorem[46,47]. Specifically, as illustrated in Fig. 3a, for any given plaquette $j$ the Berry curvature reads $\phi_j = -\text{Im}\log[\langle u_1|u_2\rangle\langle u_2|u_3\rangle\langle u_3|u_4\rangle\langle u_4|u_1\rangle]$, where $u_i$ ($i = 1\sim 4$) are four bulk states on the plaquette. Figure 3b presents our experimentally measured Berry curvature distributions for both mirror subspaces, which agree well with the theoretical predictions over a finer mesh. By accumulating the Berry curvature row by row over the discretized BZ, we further achieved the Chern numbers $C_{\pm i} = \pm 1$ for the corresponding mirror subspaces (Fig. 3c). This identifies ultimately the MCN $C_M = (C_{+i} - C_{-i})/2 = 1$, as the landmark evidence for the bulk topology of our acoustic PMCI. Note that although the measured Berry curvature deviates locally from the theoretical result, the Chern number of each mirror subspace, as a global integral, remains an integer. Mathematically, this robustness is rooted in the global topological structure of the system.

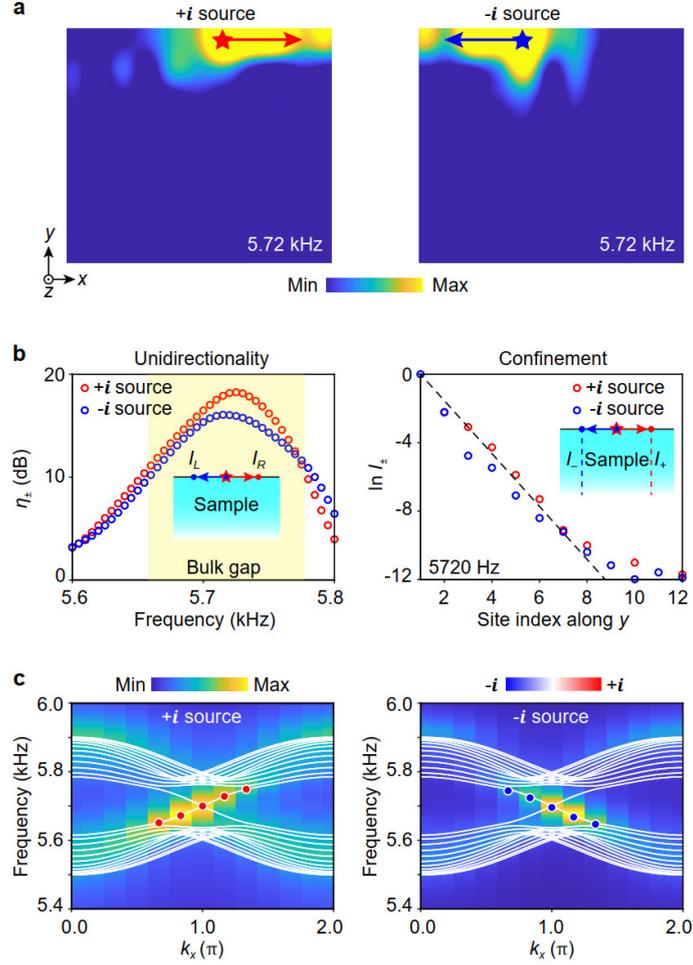

**Fig. 4 | Experimental observation of the $m_z$-locked chiral edge states. a**, Real-space visualization of the unidirectionally propagated chiral edge states confined at the top edge. The left and right panels correspond to the excitations of $+i$ and $-i$ sources, respectively. **b**, Left: spectrum response of the quantity $\eta_\pm = \pm 10\log_{10}(I_R/I_L)$, which quantitatively characterizes the unidirectionality of the $\pm i$-locked edge states in dB scale. Right: logarithmic plots for the field distributions scanned along the red ($+i$ source) and blue ($-i$ source) dashed lines, exhibiting a quantitative exponential-decay of the chiral edge states away from the top boundary. **c**, Edge spectra (color scale) selectively excited by the $\pm i$ sources, each of which features one chiral edge band as expected. Theoretical ribbon spectra (white lines) are provided for comparison, along with the measured mirror expectation values (color circles).

**Edge manifestation of the bulk topology**

Finally, we turn to the experimental characterization of the edge response to the bulk topology. According to the bulk-boundary correspondence, the bulk MCN $C_M = 1$ grants the system one pair of counter-propagated and $m_z$-locked topological edge states on each boundary[8-11]. Here we focus on the $x$-directed edge states and leave those $y$-directed ones in Supplementary Fig. 6. Note that the projective mirror symmetry is global in the bulk and preserves at the edge of a finite sample. Therefore, gapless and robust edge transport can be observed directly in a truncated PMCI, in contrast to most of the previous artificial topological insulators where specific trivial cladding crystals are required to relieve the symmetry mismatch at the boundary[48-50]. This enables more compact devices for real applications[51,52], comparing to those frequently used domain wall structures.

To directly visualize the $m_z$-resolved chiral edge states in real space, we relocated the $\pm i$ source to the middle of the top boundary, and scanned the pressure amplitude pattern over the sample. As expected in Fig. 1c, the data (exemplified at 5.72 kHz) in Fig. 4a manifest clearly the rightward and leftward propagating edge modes for the excitations of the $+i$ and $-i$ sources, respectively. (For clarity, we present only the data covering $11 \times 9$ in-plane sites.) To further quantitatively characterize the unidirectionality of the excited edge states, we extracted the sound intensities $I_R$ and $I_L$ at two unit-cells away from the source (Fig. 4b, left panel), and define the quantity $\eta_\pm = \pm 10 \log_{10}(I_R/I_L)$ for the both excitations. As shown in Fig. 4b (left panel), a unidirectionality as high as ~15 dB can be observed in both spectra within the band gap. (The difference between the results of the right and left propagations could come from some imperfections in our experimental excitations and measurements.) Together with those exponentially-decayed sound profiles away from the top boundary (Fig. 4b, right panel), we conclude the chiral and confined nature of the nontrivial edge states in each mirror subspace. The $\pm i$-locked chiral edge states that carry positive and negative group velocities can be identified further in Fig. 4c, where the Fourier spectra (color scale) are performed for the selectively-excited pressure fields scanned along the top boundary. The data agree well with the corresponding ribbon dispersions (white lines) calculated by tight-binding model. One step closer, we extracted the mirror expectation values for both excitations (Fig. 4c, color circles). The data, at an average relative error of ~2.0% with respect to $\pm i$, conclude which mirror subspaces the edge states belong to.

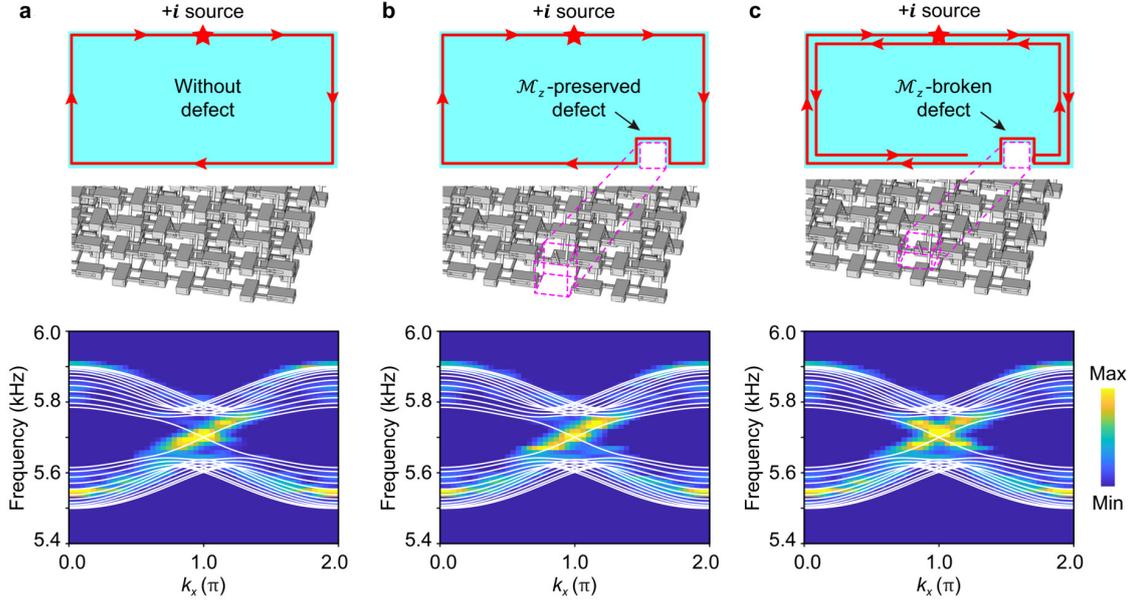

**Fig. 5 | Topological robustness of the chiral edge states against defects. a**, Numerical result for a finite sample without any defect. Top: A sketch of the edge transport excited by $+i$ source (red star). Bottom: Fourier spectrum performed for the top edge. The positive slope of the edge dispersion evidences the rightward-propagating edge states in $+i$ mirror subspace. **b**, Comparative result for a sample with a $\mathcal{M}_z$-preserved defect. The Fourier spectrum similar to **a** indicates a negligible reflection from the defect. **c**, Result for a $\mathcal{M}_z$-broken defect. The presence of the additional negative-slope dispersion shows a clear reflection from the defect.

**Robustness analysis on the topological edge states**

Finally, we present a brief discussion on the topological robustness of the $m_z$-locked chiral edge states against defects (see Fig. 5). To do this, we have simulated the transport phenomenon of the topological edge states in a finite-sized sample with a $\mathcal{M}_z$-preserved defect, in comparison to those systems without a defect and with a $\mathcal{M}_z$-broken defect. More specifically, we create defects at the lower boundary of the sample by removing acoustic cavities. When the top-layer and bottom-layer cavities are removed pairwise, we produce a defect that preserves the projective mirror symmetry $\mathcal{M}_z$ (Fig. 5b). By contrary, the symmetry is broken if we remove only a single cavity in the top layer (Fig. 5c). We simulate the associated spectrum responses for $+i$ source located in the middle of the top edge. For comparison, we present first the result for a defectless system (Fig. 5a). As shown in Fig. 5a, the Fourier spectrum performed for the top edge exhibits clearly an edge dispersion of positive slope, a manifestation of the pure excitation of the rightward edge states in $+i$ mirror subspace. As expected, similar result can be seen in Fig. 5b for the system with a $\mathcal{M}_z$-preserved defect. This means a negligible back scattering from the defect. On the contrary, for the system with a $\mathcal{M}_z$-broken defect (Fig. 5c), the presence of negative-slope edge dispersion shows a remarkable reflection from the defect.

**Conclusion and discussion**

We apply the concept of projective symmetry to construct 2D PMCIs in spinless systems, and demonstrate the unique PMCI topology in acoustic experiments. Our experimental characterization is complete and conclusive. We not only identify the distinctive mirror eigenvalue-locked bulk and edge spectra, but also directly measure the MCN through extracting the Berry curvature in momentum space. Comparing with the previous topological phases realized by projective symmetries[25-27], here the PMCI is essentially a *strong* topological phase protected by MCN. Experimentally, our protocol used for measuring topological invariants, which is efficacious even in the presence of degenerate bands, can be extended to characterize other topological phases, especially for the newly emergent Euler[53-55] and non-orientable phases[56]. Our acoustic PMCIs, featuring mirror eigenvalue-locked chiral edge states, can be used to design compact acoustic devices robust against defects and disorders, such as waveguides, resonators, and delay lines, which will lay the foundations for robust information transfer, processing, and storage. In particular, compared to the acoustic Chern insulators (which resort to circulating fluid flows to break time-reversal symmetry[57-60]), our acoustic PMCI doubles the information channels and remarkably relaxes the difficulty in practical realizations.

Note that although the MCI has been studied in condensed matter systems[8-11], the introduction of projective symmetry enriches the topology of the system. For example, the nontrivial topology in our PMCI model does not vanish after the phase transition, but transfer to a mirror $\mathbb{Z}_2$ phase. In future, one can design, fabricate, and characterize the 2D PMCI with a higher MCN, in which the integer nature of the topological index distinguishes the band topology from the other time-reversal-invariant topological phases, especially from the quantum spin Hall phase. We can also use the projective symmetry to realize 3D PMCIs, which feature gapless surface states on the mirror-invariant line of the surface BZ. Undoubtedly, the successful realization of our 2D acoustic PMCI will motivate us to explore more abundant projective symmetry-protected topological phases[56,61].

**Methods**

Our acoustic design was performed by using a commercial solver package (COMSOL Multiphysics). The photosensitive resin material used for fabricating samples was modeled as acoustically rigid in the airborne sound environment, given the extremely mismatched acoustic impedance between resin and air. Our experiments were performed for airborne sound at audible frequency. To excite the bulk and edge states, we located a pair of point-like sound sources at the top and bottom layers in the middle (top edge) of the sample, and used a needle-like microphone (B&K Type 4182) to scan the pressure information inside the cavities one by one, together with another microphone (B&K Type 4138) fixed at the source cavity for phase reference. Both the input and output signals were recorded and frequency-resolved with a multi-analyzer system (B&K Type 3560B). Through spatial Fourier transform, we obtained the bulk and edge spectra in momentum space. Importantly, to electively excite the bulk and edge states of different mirror subspaces, the pair of sound sources carry specific phase differences. Using the selectively-excited wavefunctions of each subspace (see details in *Supplementary Information*), we further computed the expectation values of the projective mirror operator, and the Berry curvature distributions for the lower bands. An integral over the first Brillouin zone gives the Chern number for each mirror subspace, which contributes the MCN ultimately.

**Data availability**

The data that support the plots and other findings of this study are available from the corresponding author upon reasonable request.

**Code availability**

Numerical simulations in this work were all performed using the 3D acoustic module of a commercial finite-element simulation software (COMSOL MULTIPHYSICS). All related codes can be built using the instructions in the Method section.

**Acknowledgements**

This project is supported by the National Natural Science Foundation of China (Grant No. 11890701, 12374418, and 12104346), and the Young Top-Notch Talent for Ten Thousand Talent Program (2019-2022).

**Author contributions**

C.Q. conceived the idea and supervised the project. T.L. developed the theory and did the simulations. T.L. designed and performed the experiments under the help of L.L. and Q.Z. T.L. and C.Q. analyzed the data and wrote the manuscript. All authors contributed to scientific discussions of the manuscript.

**Competing interests**

The authors declare no competing interests.

**Author information**

Correspondence and requests for materials should be addressed to C.Q. (cyqiu@whu.edu.cn).